\documentclass[aps,prl,reprint,groupedaddress,showpacs,longbibliography]{revtex4-2}

\usepackage{amssymb,amsmath} 
\usepackage{graphicx}

\newcommand{\mi}{\mathrm{i}}
\usepackage[colorlinks=true,  urlcolor=blue,  linkcolor=blue,  citecolor=blue]{hyperref}

\begin{document}

\title{Realization of a Townes Soliton in a Two-Component Planar Bose Gas}

\author{B. Bakkali-Hassani$^{1}$, C. Maury$^{1}$, Y.-Q. Zou$^{1}$, \'E. Le Cerf$^{1}$, R. Saint-Jalm$^{3}$, P.C.M. Castilho$^{2}$, S. Nascimbene$^{1}$, J. Dalibard$^{1}$, J. Beugnon$^{1}$}

\affiliation{$^{1}$Laboratoire Kastler Brossel,  Coll\`ege de France, CNRS, ENS-PSL University, Sorbonne Universit\'e, 11 Place Marcelin Berthelot, 75005 Paris, France}
\affiliation{$^{2}$Instituto de F\' isica de S\~ao Carlos, Universidade de S\~ao Paulo, CP 369, 13560-970 S\~ao Carlos, Brazil}
\affiliation{$^{3}$Department of Physics, Ludwig-Maximilians-Universit\"at M\"unchen, Schellingstr. 4, D-80799 M\"unchen, Germany}

\date{\today}

\begin{abstract}
Most experimental observations of solitons are limited to one-dimensional (1D) situations, where they are naturally stable. For instance, in 1D cold Bose gases, they exist for any attractive interaction strength $g$ and particle  number $N$. By contrast, in two dimensions, solitons appear only for discrete values of $gN$, the so-called Townes soliton being the most celebrated example. Here, we use a two-component Bose gas to prepare deterministically such a soliton: Starting from a uniform bath of atoms in a given internal state, we imprint the soliton wave function using an optical transfer to another state. We explore various interaction strengths, atom numbers and sizes, and confirm the existence of a solitonic behaviour for a specific value of $gN$ and arbitrary sizes, a hallmark of scale invariance.
\end{abstract}

\maketitle

Solitary waves are encountered in a broad range of fields, including photonics, hydrodynamics, condensed matter and high-energy physics \cite{Dauxois06}. The solitonic behavior usually originates from the balance between the  tendency to expansion of a wave packet in a dispersive medium and a non-linear contracting effect. For a real or complex field $\phi$ in dimension $D$, a paradigm example is provided by the energy functional 
\begin{eqnarray}
E[\phi]=\frac{1}{2} \int \mathrm{d}^Dr \,\left[ | \boldsymbol{\nabla}\phi(\boldsymbol{r})|^2-G|\phi(\boldsymbol{r})|^4 \right], \label{eq_E}
\end{eqnarray}
where $G$ is a real positive parameter and $\phi$ is normalized to unity ($\int \mathrm{d}^Dr \,|\phi|^2=1$). This energy functional is relevant for describing the propagation of an intense laser beam in a non-linear cubic medium, where diffraction and self-focusing compete. It is also used to model the evolution of coherent matter waves when the kinetic energy contribution competes with attractive interactions.

Experimentally, most studies concentrate on effective one-dimensional situations, as solitons in higher dimensions are more prone to instabilities and then much more challenging to investigate experimentally \cite{Kartashov19}. This can be understood from a simple scaling analysis of $E[\phi]$, assuming a wave packet $\phi$ of size $\ell$:
\begin{equation}
E(\ell)\sim \frac{1}{\ell^2}-\frac{G}{\ell^D}.
\label{eq:E_ell}
\end{equation}
For $D=1$, this leads to a stable minimum for $\ell^*\sim 1/G$ whereas for $D=3$, the extremum obtained for $\ell^* \sim G$ is dynamically unstable.

The case $D=2$ is of specific interest because the two terms of the estimate of Eq.\,(\ref{eq:E_ell}) scale as $1/\ell^2$. The existence of a localized wave packet minimizing $E[\phi]$ (or more generally letting it stationary) can thus be achieved only for discrete values of $G$. The Townes soliton, introduced in Ref.\,\cite{Chiao64}, is a celebrated example of such a stationary state. It is the real, nodeless and axially symmetric solution of the two-dimensional (2D) Gross-Pitaevskii or non-linear Schr\"odinger equation (NLSE) obtained by imposing $\delta E[\phi]=0$:
\begin{eqnarray}
-\frac{1}{2} \nabla^2 \phi(\boldsymbol{r})- G \phi^3(\boldsymbol{r})=\mu\, \phi(\boldsymbol{r}), \label{eq:NLSEeff}
\end{eqnarray}
where  the chemical potential $\mu$ can take any negative value. These solutions, which have zero energy ($E[\phi]=0$), exist only for $G=G_T\approx5.85$.  Scale-invariance of the 2D-NLSE \cite{Pitaevskii97a} provides a relation between them: if $\phi(\boldsymbol{r})$ is solution of Eq.\,\eqref{eq:NLSEeff} for a given $\mu$, then for any real $\lambda$, $\lambda\phi(\lambda \boldsymbol{r})$ is also a normalized solution with a rescaled $\mu$, still with zero energy. For $G<G_T$, there are no stationary localized solution of Eq.\,\eqref{eq:NLSEeff}, while for $G>G_T$, one can find localized functions $\phi$ with an arbitrarily large and negative energy $E[\phi]$.

Townes solitons have been mostly investigated in non-linear optics where the cubic term in Eq.\,\eqref{eq:NLSEeff} corresponds to a  Kerr non-linearity that induces self-focusing for intense enough beams \cite{Bjorkholm74,Barthelemy85,Moll03}. The soliton solution then corresponds to a specific optical power. Numerous strategies have  been developed to stabilize the Townes soliton and to observe some features of solitonic propagation in various optical settings \cite{Torruellas95,Falcao13,Reyna14,Duree93,Pasquazi07,Fleischer03,Cerda-Mendez13}. Ultracold gases are another well-known platform to investigate soliton physics in 1D \cite{Burger99,Denschlag00,Khaykovich02,Strecker02,Eiermann04}. The 2D case has been recently explored in Ref.\,\cite{Chen20} using a quench of the interaction strength to negative values. The formation of multiple stationary wave packets was observed, where the size of a wave packet was fixed by the out-of-equilibrium dynamics.

In this Letter, we report the first deterministic realization of a 2D matter-wave Townes soliton. We demonstrate, at a given interaction strength, the existence of stationary states for a well-defined atom number and for the specific ``Townes profile" of the density distribution. We also show that this behavior is independent of the wave packet size, hence confirming scale-invariance. To produce this soliton, we use a novel approach based on a two-component Bose gas for which  the equilibrium state of one minority component immersed in a bath defined by the other component is well-described, in the weak depletion regime, by an effective single-component NLSE with cubic non-linearity.

We consider atoms of mass $m$ in states $|1\rangle$ and $|2\rangle$ with repulsive contact interactions. The intracomponent ($\tilde g_{11}$, $\tilde g_{22})$ and the intercomponent ($\tilde g_{12}$) interaction parameters are thus all positive. Such a mixture is well described in the zero-temperature limit by two coupled NLSEs. In the weak depletion regime, one can assume that the dynamics of the dense bath of atoms in state $|1\rangle$ with density $n_1$ occurs on a short time scale ($\propto m/(\hbar^2 \tilde g_{11} n_1)$) compared to the minority component dynamics. The bath is then always at equilibrium on the time scale of the evolution of the minority  component. 
The equilibrium state $\phi(\boldsymbol{r})$ for $N$ particles in state $|2\rangle$ then satisfies the single component NLSE given in Eq.\,(\ref{eq:NLSEeff}) with
\begin{eqnarray}
G=-N\tilde g_e,\quad\tilde g_e=\tilde g_{22}-\frac{\tilde g_{12}^2}{\tilde g_{11}}. \label{eq:ge}
\end{eqnarray}
The effective interaction parameter $\tilde g_e$ corresponds to a dressing of the interactions for component $|2\rangle$ by the bath  \cite{Pethick08}. In this  limit, the dynamics of the particles in state $|2\rangle$ remains scale-invariant since the characteristic length of the bath, \emph{i.e.} its healing length, does not play any role. We discuss at the end of this Letter possible deviations from this limit.

The experiments described in this Letter focus on the case of $^{87}$Rb atoms in their electronic ground level where all interaction parameters $\tilde g_{ij}$ are close to each other within a few percent. Here we use the  $|1\rangle=|F=1,m_F=0\rangle$ and  $|2\rangle=|F=2,m_F=0\rangle$ states for which we have $\tilde g_e<0$. The negative value of $\tilde g_e$ implies that the effective dynamics of the minority component is akin to the one of a gas with attractive interactions, as required for observing Townes solitons. The condition for effective attractive interactions is also equivalent to the immiscibility criterion for the two components \cite{Timmermans98}. The atom number $N_T$ corresponding to the Townes soliton for a cloud with interaction parameter $\tilde g_e$ is then given by $N_T = G_T / |\tilde g_e|$.

\begin{figure}
\centering
\includegraphics[width=8.6cm]{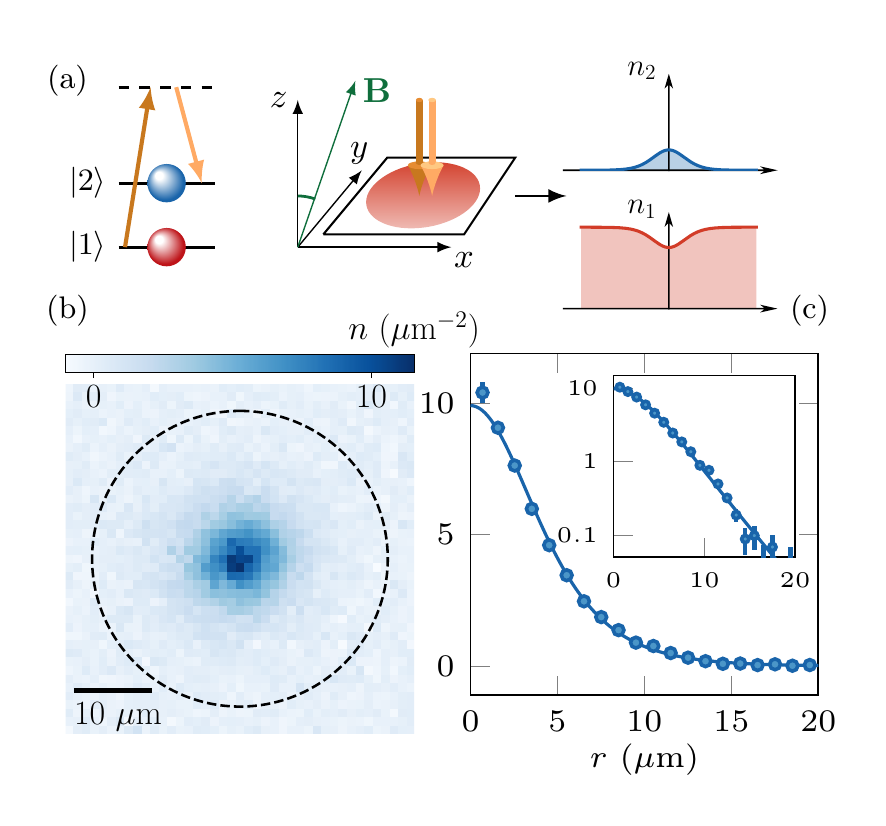}
\caption{(a) Schematics of the experiment. We create a disk-shaped planar Bose gas in the $xy$ plane in state $|1\rangle$. At time $t=0$ we pulse a pair of copropagating Raman beams which transfer in a spatially-resolved way a small fraction of the atoms from state $|1\rangle$ to state $|2\rangle$. An example of density distribution $n_2 \equiv n$ in state $|2\rangle$ obtained when preparing a Townes profile is shown in (b), the dashed line indicates the edge of the bath of atoms in state $|1\rangle$. Its radial profile is reported in (c) as blue dots, together with its fit to a Townes density profile (solid line). The inset displays the same data in semilog scale highlighting the approximately exponential tails of the Townes profile.}
\label{fig1}
\end{figure}

Our experimental study of Townes solitons starts with the preparation of a uniform two-dimensional Bose gas of $^{87}$Rb atoms in state $|1\rangle$, as detailed in Refs.\,\cite{Ville17,SaintJalm19}. Atoms are confined in a circularly-shaped box potential in the horizontal plane and they occupy the ground-state of an approximately harmonic potential along the vertical direction. The typical cloud temperature is $<20\,$nK and the column density is set around $100$ atoms$/\mu$m$^2$. In this regime, the gas is well described by a 2D cubic NLSE. An external magnetic field of $0.7$\,G with tunable orientation is applied.

At time $t=0$, we create a custom-shaped wave packet of atoms in component $|2\rangle$ immersed in a bath of atoms in component $|1\rangle$, as shown in Fig.\,\ref{fig1}a.  This is achieved by  transferring, in a spatially-resolved way, a controlled fraction of atoms into  $|2\rangle$ thanks to a two-photon Raman transition, which keeps the total density constant. We use two colinear laser beams so as not to impart any significant momentum to the transferred atoms. The in plane intensity profile of these beams is shaped by a spatial light modulator, which allows us to design arbitrary intensity patterns on the atomic cloud with about 1\,$\mu m$ spatial resolution \cite{REFSM,Zou21}. We show  in Fig.\,\ref{fig1}b and  \ref{fig1}c an example realization of a Townes profile, i.e. a density distribution $n(\boldsymbol{r})$ proportional to $|\phi(\boldsymbol{r})|^2$, where $\phi$ is obtained by a numerical resolution of Eq.\,\eqref{eq:NLSEeff}. It shows an excellent control of the density distribution of component $|2\rangle$ over more than two decades in density.  After imprinting a Townes profile of given amplitude and width, we let the system evolve and we measure the in situ density distribution via absorption imaging. All the profiles studied here are initially in the weak depletion regime, where the density $n$ does not exceed 20$\%$ of the bath density. We restrict the time evolution to durations short enough to limit the amount of losses in state $|2\rangle$, essentially due to hyperfine relaxation, to typically $\lesssim 10\%$.

\begin{figure}[t!!]
\centering
\includegraphics[width=8.6cm]{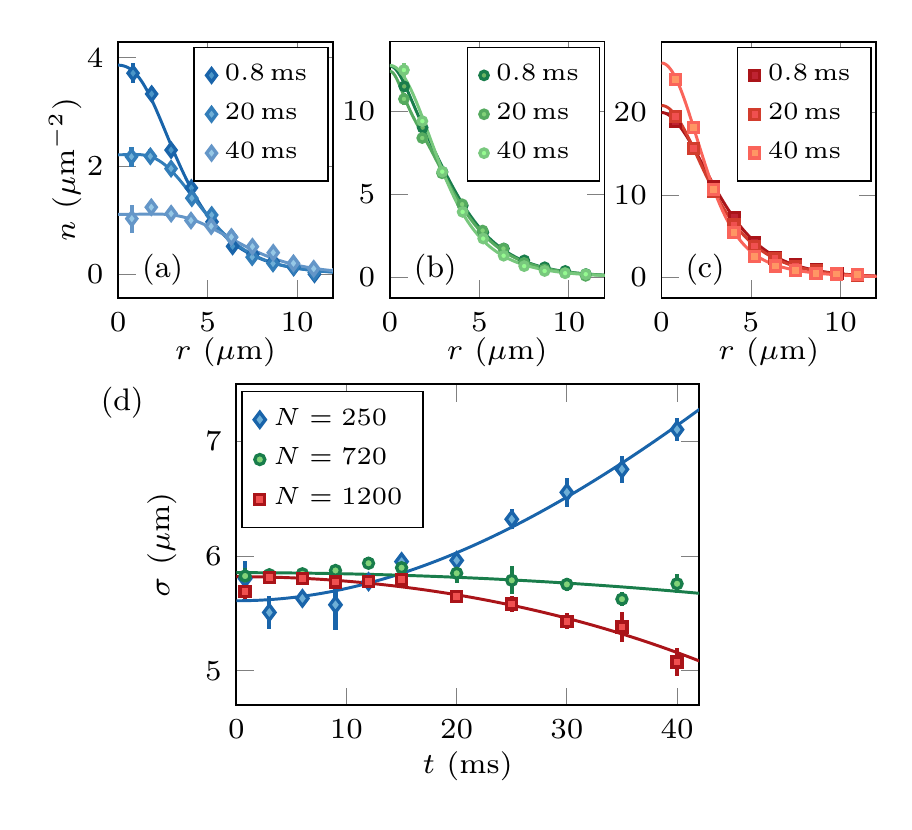}
\caption{ (a-c) Radial profiles at different times for  imprinted Townes profiles with (a) $N=250 (40)$, (b)  $N=720 (20)$ and (c) $N=1200 (50)$ atoms. Initial rms sizes are similar and the magnetic field is perpendicular to the atomic plane. The solid lines are fits to the data. (d) Time evolution of the fitted rms size for the same three configurations. The solid lines are a fit to the data with Eq.\,\eqref{eq:defgamma}.}
\label{fig2}
\end{figure}

The two states used in this work are characterized by their $s$-wave scattering lengths $a_{11}=100.9\,a_0$, $a_{22}-a_{11}=-6.0\,a_0$ and $a_{12}-a_{11}=-2.0\,a_0$ \cite{Altin11}, where $a_0$ is the Bohr radius. Thanks to the existence of magnetic dipole-dipole interactions in  a mixture of the two components, the value of $a_{12}$ can be shifted, for a 2D cloud, by an amount varying from $-0.7\,a_0$ to $+1.4\,a_0$ by changing the angle of the applied magnetic field with respect to the atomic plane \cite{Zou20a}. In all cases, we have $a_{22}-a_{12}^2/a_{11}<0$ and thus a similar inequality for the interaction parameters defined as $\tilde g_{ij}=\sqrt{8\pi} a_{ij}/ \ell_z$ for $i,j=1,2$, where $\ell_z=\sqrt{\hbar/m\omega_z}$ is the harmonic oscillator length associated to the confinement along the vertical direction of frequency $\omega_z$. Here, we have $\tilde g_{11} =0.16(1)$.

In Fig.\,\ref{fig2}a-c, we show, for three different atom numbers, the measured time evolution of a Townes profile with a root-mean-square (rms) size at time $t=0$ given by $\sigma_0=5.7\,\mu$m and the external magnetic field perpendicular to the atomic plane ($\tilde g_e \approx -7.6\times 10^{-3}$). We observe an almost stationary time evolution for $N=720(20)$ whereas the central density of the cloud decreases for $N=250(40)$ and increases for $N=1200(50)$. More quantitatively, for each time $t$ we extract the rms size  $\sigma(t)$  of the cloud (see \cite{REFSM} for details) and we study its time evolution, as shown in Fig.\,\ref{fig2}d.

We analyze these data using the variance identity (or virial theorem), which provides the time evolution of the rms size of the density profile for the 2D NLSE \cite{Pitaevskii97a}
\begin{eqnarray}
\frac{\mathrm d^2\sigma^2}{\mathrm dt^2} = \frac{4E}{m}, \label{eq:virial}
\end{eqnarray}
where $E$ is the total (kinetic+interaction) energy per particle. We thus fit the time evolution of $\sigma$ to the function resulting from the integration of Eq.\,\eqref{eq:virial}
\begin{eqnarray}
\sigma^2(t)=\sigma_0^2 + \left(\frac{\hbar}{m \sigma_0}\right)^2\,\gamma\, t^2,\label{eq:defgamma}
\end{eqnarray}
where we assumed that the imprinted state is a real wave function and thus $\mathrm{d}\sigma/\mathrm{d}t =0$ at $t=0$. For the Townes profile, one can show that the explicit expressions of the kinetic and interaction energy integrals lead to $\gamma=\alpha(1-N/N_T)$, where $\alpha\approx 1.19$ is determined numerically (note that $\gamma=1$ for a non-interacting Gaussian wave packet). 

\begin{figure}
\centering
\includegraphics[width=8.6cm]{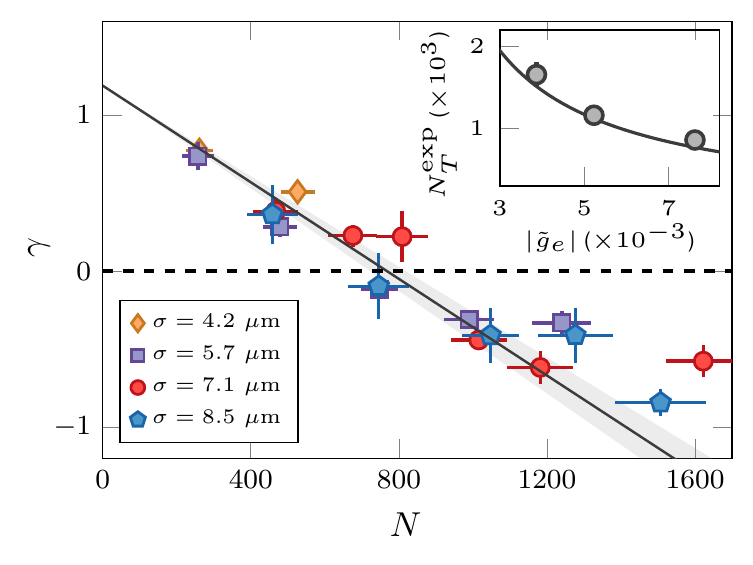}
\caption{Expansion coefficient as a function of the atom number of the imprinted wave packet for a magnetic field perpendicular to the atomic plane. All data for different initial sizes  collapse onto a single curve. The solid line is the theory prediction computed for $\tilde g_e = -7.6\times 10^{-3}$ without any adjustable parameter. The shaded area around this line represents our estimated uncertainty on the calibration of $\tilde g_e$. (Inset) Variation of the experimentally determined stationary atom number $N_T^{\rm exp}$ for different values of $\tilde g_e$. The stationary atom number is determined from a linear fit of the various $\gamma(N)$ curves \cite{REFSM}. The solid line is the prediction $N_T=G_T/|\tilde g_e|$.}
\label{fig3}
\end{figure}

We report in Fig.\,\ref{fig3} the fitted expansion coefficient $\gamma$ as a function of the atom number $N$ for different values of the initial size $\sigma_0$. All data collapse onto a single curve $\gamma(N)$ which experimentally confirms the scale-invariance of the system. The stationary state, $\gamma=0$, is obtained for $N_T^{\rm exp}=790(40)$ (determined with a linear fit). We also show as a solid line the prediction $\gamma=\alpha(1-N/N_T)$, where $N_T=770(50)$ is fixed by the independently estimated value of $\tilde g_e$ \footnote{The reported uncertainties on the measured atom number are associated to the statistical variations of the cloud over the different repetitions of the experiment. Systematic errors on the atom number calibration are estimated to be on the order of 10\%. The determination of $N_T$ is sensitive to the knowledge of the scattering length differences. A variation of these two differences by $0.1\,a_0$ corresponds to a variation of $N_T$ by $\approx$ 50 atoms for our experimental parameters.}. It shows a very good agreement for lower values of $N$. The small deviation at large $N$ is likely due to the larger density of the minority component wave packet, which leads to increased losses and deviation from the low depletion regime.

\begin{figure}[t!]
\centering
\hskip-15pt\includegraphics[width=9cm]{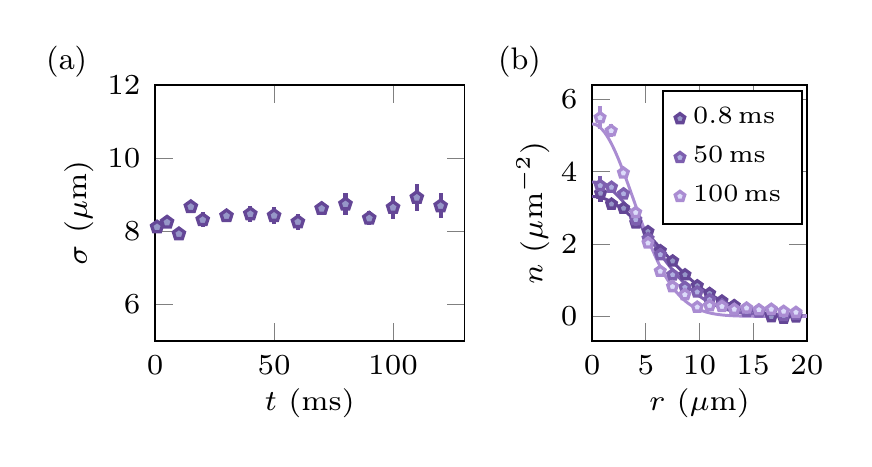}
\caption{Time evolution of a Gaussian profile with $N \sim 800 \sim N_G$. (a) The chosen atom number corresponds to a zero energy state as shown by the almost stationary rms size. However, the density profile shown in (b) evolves with time in contrast to the Townes profile shown in Fig.\,\ref{fig2}b.}
\label{fig4}
\end{figure}

The relevant quantity to determine the behavior of the imprinted wave packet is $|\tilde g_e| N$ that should be compared to $G_T=5.85$. We show in the inset of Fig.\,\ref{fig3} the measured variation of $N_T^{\rm exp}$ when varying the orientation of the applied magnetic field with respect to the atomic plane and hence the interspecies scattering length. We confirm the prediction $N_T = G_T/| \tilde g_e|$ with $\tilde g_e$ varying from $-3.9$ to $-7.6$ $\times 10^{-3}$ \cite{Zou20a,REFSM}.

For an arbitrary density profile there always exists an atom number such that the energy of Eq.\,$\eqref{eq_E}$ is zero and hence, from Eq.\,\eqref{eq:virial}, the rms size is stationary. Of course, this is not sufficient to achieve a fully stationary profile. We illustrate this point in Fig.\,\ref{fig4} for the case of an initial Gaussian profile, for which a  zero energy is obtained for $N_G=2\pi/|\tilde g_e|$ \cite{Fibich00}. We check in Fig.\,\ref{fig4}a that this number leads to a stationary rms. However,  the observed density distribution is clearly not stationary as shown in Fig.\,\ref{fig4}b.

Our approach using a two-component gas raises new specific questions. For instance, for a wave packet with  large enough $G$,  the central density can diverge at a finite time in the single-component case whereas such a collapsing behaviour cannot occur in the two-component case with repulsive interactions (all $\tilde g_{ij}>0$). Indeed, as the minority component density becomes comparable to  the bath one, the bath  brings a new length scale to the effective one-component description,  thus breaking scale invariance. Let us briefly discuss this problem in the case of close interaction parameters $\tilde g_{ij}$, which is  relevant for the two states of $^{87}$Rb used here. In this limit, the coupled NLSEs describing the binary system can be simplified into a single one describing the equilibrium state of component $|2\rangle$, without requiring a weak depletion approximation \cite{REFSM}. Introducing the bath density  $n_\infty$, we expand this single-component equation at first order $\phi^2/n_\infty$ and obtain
\begin{eqnarray}
 - \frac{1}{2}\nabla^2 \phi-G \phi^3-\frac{N}{4}\frac{(\nabla^2 \phi^2)}{n_\infty}\phi= \mu\, \phi. \label{eq:NLSEeff3}
\end{eqnarray}
We recover Eq.\,(\ref{eq:NLSEeff}) with the same effective interaction parameter $G=-N\tilde g_e$, but with an additional stabilizing term which breaks scale-invariance. The influence of this term was investigated in a different context in Ref.\,\cite{Rosanov02}. Contrary to the case of the cubic equation, it leads for  any atom number $N>N_T$ to a localized ground state solution  with a well-defined size $\sigma_N$. We checked that for all data reported in Fig.\,\ref{fig3} the shift of the stationary atom number due to the additional stabilizing term of Eq.\,\eqref{eq:NLSEeff3} remains small ($\lesssim 10\%$) \cite{REFSM}. 

Scale invariance is also broken in the one-component case when one regularizes the contact potential that leads to the interaction energy term $-G\int|\phi|^4$ in Eq.\,(\ref{eq_E}) \cite{Hammer04,Lee06,Bazak18}. Such a regularization is not required as long as one restricts to the classical field approach of Eqs.\,(\ref{eq_E},\ref{eq:ge}), valid for $|\tilde g_e| \ll 1$  \cite{Svistunov:2015}, but it becomes compulsory for larger $|\tilde g_e|$ where a quantum treatment of atomic interactions is in order. After regularization, the interaction strength $\tilde g_e$ becomes a running coupling constant. Then, there exists a stable  solution of size $\sigma_N$ and energy $E_N$ for any value of the atom number $N$ with the geometric scaling $\sigma_{N+1}/\sigma_N \eqsim 0.34$ \cite{Hammer04}. In practice, the predicted value for $\sigma_N$ is physically reasonable only for $|N-N_T| \sim$ few units.  Morever, for $|\tilde g_e|\lll1$, as explored here, the typical  evolution time scale of a  $N$-particle state with a Townes profile of size $\sigma$ slightly different from $\sigma_N$ will be prohibitively long \cite{REFSM}. In the case $|\tilde g_e| \sim 1$, a realistic droplet size would be achieved for only a few atoms  and one could observe the predicted scaling of $\sigma_N$ with $N$.

It is also  interesting to put our  work in perspective with the physics of quantum droplets \cite{Bulgac02,Petrov15} or mixed bubbles \cite{Naidon20}, which has recently attracted great interest. Such droplets have been observed   in  1D or 3D geometries \cite{Schmitt16,Ferrier16,Chomaz16,Cabrera18,Semeghini18,Cheiney18}. Their formation results from the competition between a tunable mean-field attractive term and a beyond-mean field repulsive term. The scaling of the two terms with density is different and leads to a stable equilibrium with a droplet size that depends on the particle number.  In this Letter, the observed 2D solitons are purely mean-field objects resulting from the balance between effective attractive interactions and kinetic energy.

To summarize, we have presented a new platform to explore the physics of solitons in two dimensions.  Higher order solutions of the 2D NLSE, with nodes in the density profile \cite{Haus66,Zakharov71} or vortex solitons \cite{Kruglov85}, can also be investigated with similar methods. Another natural extension consists in printing solitons with a well-defined momentum imparted by the two-photon Raman transfer. Propagation, interaction or fusion of solitons could then be explored \cite{Stegeman99,Cornish06,Nguyen14,Ferioli19}. Additionally, whereas we focused here on the equilibrium solution at zero temperature, it will be interesting to study the elementary excitations of these solitons \cite{Malkin91}, as well as the role of finite temperature on the dynamical behavior of these objects \cite{Sinha06}.

\begin{acknowledgments}
This work is supported by ERC (Synergy UQUAM), European Union's Horizon 2020 Programme (QuantERA NAQUAS project) and the ANR-18-CE30-0010 grant. We thank D. Petrov for fruitful discussions and G. Chauveau for his participation to the final stage of the project.
\end{acknowledgments}

\bibliography{Townes_bib}

\appendix

\clearpage

\section{SUPPLEMENTAL MATERIAL}

\section{1 - Raman beam shaping}

We describe the procedure for preparing a two-component gas with a specific spin pattern, as reported in the main text. We start from a homogeneous sample of atoms in state $|1 \rangle$ filling a disk-shaped box potential of radius $R = 20\,\mu$m, with a 2D-density defined as $n_{\infty}$. Atoms are transferred from state $|1 \rangle$ to state $|2 \rangle$ using a pair of co-propagating Raman beams along the $\hat{z}$-direction, the two beams having the same waist $w \sim 40\,\mu$m. The frequency difference between the two beams is resonant with the hyperfine energy splitting of $6.8$\,GHz between the two states. In addition, the wavelength of each beam is set to $\lambda \simeq 789.9$\,nm, in between the D1 and D2 lines of $^{87}$Rb. This allows us to cancel the scalar light-shift induced by the Raman beams which could, because of  intensity gradients, print a non-uniform phase on the atomic states over the cloud size. The Raman pulse duration is short enough ($< 25\,\mu$s for all data studied in the main text) so that no dynamics occur during the transfer. 

Before reaching the atomic plane, the Raman beams reflect on a DMD (DLP7000 from Texas Instruments interfaced by Vialux GmbH) which we use as an intensity modulator to tune the intensity and hence the local Rabi frequency of the Raman beams driving the atomic transition. Despite the fact that such a modulator displays a binary image (``black or white"), we can create a grey-level image on the atoms by averaging the contribution of many pixels over a size of $1\,\mu$m, which corresponds to both our typical optical resolution and the effective pixel size in the atomic plane of the camera used to image the cloud.
The protocol to create such spin patterns is  based on an iterative algorithm which minimizes the difference between the measured spin distribution and the targeted one and is discussed in more detail in Ref.\,\cite{Zou21}.

\section{2 - Effective single-component description}

We present the derivation of an effective single-component description of our two-component system, focusing on the ground state wavefunction. The atomic mixture is described by two coupled non-linear Schr\"odinger equations (NLSEs)
\begin{align}
\tilde\mu_1\phi_1&=-\frac{1}{2}\nabla^2\phi_1+(\tilde g_{11}n_1+\tilde g_{12}n_2)\phi_1,\label{eq1}\\
\tilde\mu_2\phi_2&=-\frac{1}{2}\nabla^2\phi_2+(\tilde g_{12}n_1+\tilde g_{22}n_2)\phi_2,\label{eq2}
\end{align}
where $n_i$ is the atomic density for the spin state $i$. We also introduce the reduced chemical potentials $\tilde\mu_i=m\mu_i/\hbar^2$. We are interested in localized wavefunctions for the component $|2\rangle$ immersed in a bath of atoms in state $|1\rangle$ extending to infinity. Therefore, the chemical potential $\mu_1$ for the component $|1\rangle$ equals the mean field energy shift $g_{11}n_\infty$ at the asymptotic density $n_\infty$. 

The effective single-component description relies on the vicinity of the interaction coupling constants, i.e.
\begin{equation}
 \frac{|\tilde g_{12}-\tilde g_{11}|}{\tilde g_{11}},\quad \frac{|\tilde g_{22}-\tilde g_{11}|}{\tilde g_{11}}\ll 1, \label{eq:deltagij}
\end{equation}
which allows one to simplify the NLSE at lowest order in these small parameters. In this situation, we expect the low-energy dynamics to be dominated by spin waves, such that the total density $n_1+n_2=n_\infty+\delta n$ is weakly perturbed, with an excess density $\delta n$ satisfying $|\delta n|\ll n_\infty$. At low energy, the relevant spatial variations occur on the scale of the spin healing length \cite{Timmermans98}, which largely exceeds the bath healing length $\xi = 1/\sqrt{\tilde g_{11}n_\infty}$. Therefore,  the Laplacian operator can itself be considered of order one in the small parameters defined in Eq.\,\eqref{eq:deltagij}, such that the term $\nabla^2\phi_1$ in Eq.\,\eqref{eq1} can be replaced, at order one, by $\nabla^2\sqrt{n_\infty-n_2}$ (assuming a real-valued wavefunction). This approximation allows one to express the excess density $\delta n$ in terms of the second component only, as
\begin{equation}
\tilde  g_{11}\delta n=\frac{\nabla^2\sqrt{n_\infty-n_2}}{2\sqrt{n_\infty-n_2}}+(\tilde  g_{11}-\tilde  g_{12})n_2.
\end{equation}
Inserting this expression in Eq.\,\eqref{eq2}, we obtain an effective single-component equation for component $|2\rangle$. As we focus only on component $|2\rangle$ hereafter, we drop the index $2$ ($\phi_2,n_2,\tilde\mu_2 \rightarrow \phi,n,\tilde\mu$) and write the effective equation
\begin{equation}
 \tilde\mu\phi=\tilde g_{12}n_\infty\phi-\frac{1}{2}\nabla^2\phi+\tilde g_e n \phi+\frac{\nabla^2\sqrt{n_\infty-n}}{2\sqrt{n_\infty-n}}\phi,\label{NLSE_eff}
\end{equation}
where we introduce the effective coupling constant
\begin{equation}
\tilde  g_e=\tilde g_{22}-\frac{\tilde g_{12}^2}{\tilde g_{11}}.
\end{equation}
The term $\tilde g_{12}n_\infty\phi$ corresponds to the interaction energy cost for adding a single particle of component $|2\rangle$ into the bath. Such a global energy shift plays no role in the following, and we absorb it in the chemical potential hereafter.
Eq.\,\eqref{NLSE_eff} is a non-linear Schr\"odinger equation with two non-linear terms. The term $\tilde g_e n \phi$ is a standard cubic nonlinearity, corresponding to an effective system of bosonic particles with contact interactions and coupling constant $\tilde g_e$ \cite{Pethick08}. The second term is more complex and plays a significant role when the density $n$ becomes comparable to the asymptotic bath density $n_\infty$. To be more precise, one can expand, in  the limit of large bath density, Eq.\,\eqref{NLSE_eff} in powers of the depletion $n/n_\infty$. At minimal order we obtain the NLSE used in the main text
\begin{equation}
 \tilde\mu\phi=-\frac{1}{2}\nabla^2\phi+\tilde g_{e}n\phi,\label{NLSE_eff_2}
\end{equation}
with the coupling constant $\tilde g_e$. 

In the case $\tilde g_e < 0$ relevant for our experiments, this equation has, for each negative value of the chemical potential, a localized stationary solution -- the so-called Townes soliton -- that can be written as 
\begin{equation}
\phi_\ell(r) = \frac{1}{\ell\sqrt{G_T}} R(r/\ell), \label{eq:townes_family}
\end{equation}
where we introduce the length $\ell=1/\sqrt{|\tilde\mu|}$ and 
$R$ is the zero-node solution of the differential equation
\begin{equation}
 \left(\frac{1}{2}\nabla^2+R^2-1\right)R=0.\label{eq:defR}
\end{equation}
This function is normalized to the value $G_T = \int \mathrm{d}^2r\, R^2(r) \simeq 5.850$.
The wave functions $\phi_\ell(r)$ correspond to zero-energy states that have the same atom number, equal to 
\begin{equation}
 N = N_T = \frac{G_T}{|\tilde g_e|}.
\end{equation}
The self-similar nature of this family of solutions reflects the scale invariance of the NLSE in two dimensions given in Eq.\,\eqref{NLSE_eff_2}.

At next order in the perturbation, we obtain the equation
\begin{equation}
\tilde\mu\phi=-\frac{1}{2}\nabla^2\phi+\tilde g_e n \phi-\frac{\nabla^2 n}{4n_\infty}\phi.\label{NLSE_eff_3}
\end{equation}
The additional term, which was considered in \cite{Rosanov02}, can be viewed as a weakly non-local interaction. Since it involves an explicit length scale $1/\sqrt{n_\infty}$, it breaks scale invariance, and we no longer expect self-similarity between stationary states. In a linear perturbative treatment, the stationary state is written as a weakly deformed Townes soliton
\begin{equation}
\phi(r) \propto \frac{1}{\ell} R(r/\ell) + \frac{1}{2 n_\infty \ell^3}R_2(r/\ell),
\end{equation}
 where $R$ is defined in Eq.\,\eqref{eq:defR} and $R_2$ is the solution of
\begin{equation}
\left(\frac{1}{2}\nabla^2+3R^2-1\right)R_2=-\frac{1}{2} R \, \nabla^2 R^2 .
\end{equation}
The atom number contained in the perturbed state is always larger than $N_T$ and is pertubatively given by
\begin{equation}
 N\simeq N_T\left(1 + 0.23\frac{N_T}{n_\infty \ell^2}\right). \label{eq:Rosanov}
\end{equation}
This prediction is in good agreement with the results of numerical simulations described in the following Section.

\section{3 - Beyond the weak depletion limit: simulations}

We explore here the ground-state properties of our two-component system beyond the weak depletion regime. We compare the different approaches introduced in Section\,2 of these Supplemental materials: 
\begin{itemize}
\item[--]   The two coupled NLSEs given by Eqs.\,\eqref{eq1}-\eqref{eq2}.
\item[--]   The single component effective equation given in Eq.\eqref{NLSE_eff}, valid for arbitrary depletion and close $g_{ij}$'s.
\item[--]  The low depletion limit of the previous equation given by Eq.\,\eqref{NLSE_eff_3} including the first order correction to the scale invariant attractive NLSE.
\end{itemize}

\begin{figure}
\centering
\includegraphics[width=8.2cm]{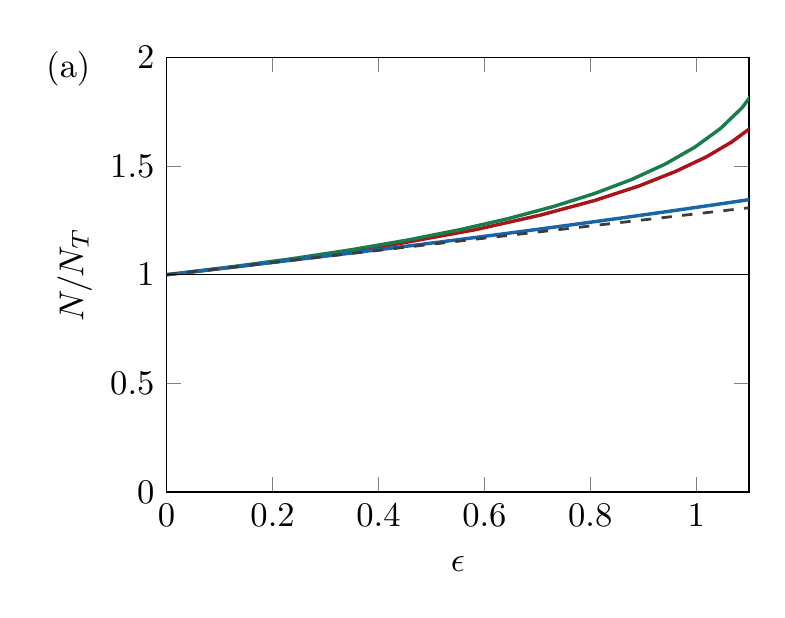}
\includegraphics[width=8.2cm]{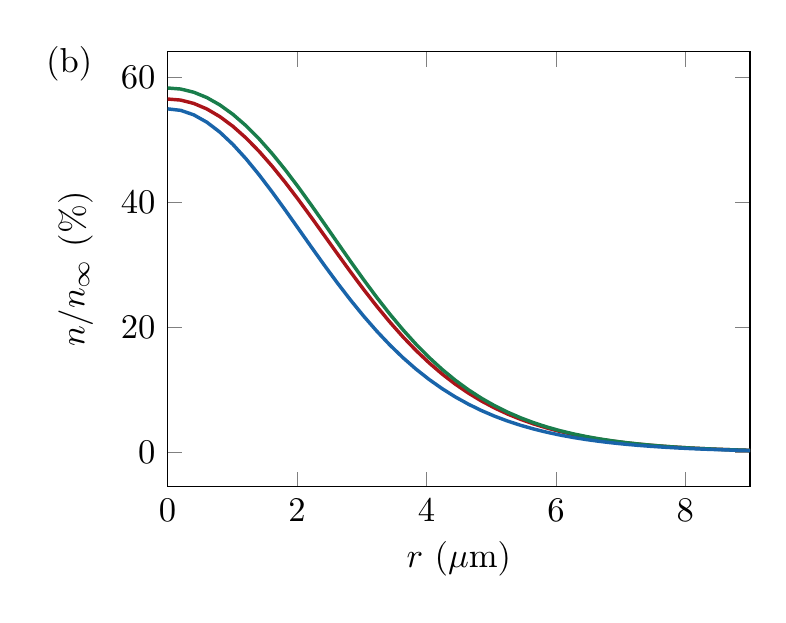}
\caption{Numerical study of the ground state for different models. (a) Deviation of the ground state atom number $N$ with respect to $N_T$ as a function of the depletion parameter $\epsilon$ for different models: two-component NLSE (red) with $\tilde g_{11}=0.16$ and $(\tilde g_{12}, \tilde g_{22}) = (0.98, 0.94) \, \tilde g_{11}$, effective one-component NLSE (green), weak depletion expansion of the effective one-component NLSE (blue), analytical prediction of Eq.\,\eqref{eq:Rosanov_2} (dashed black). (b) Radial profiles for $\epsilon=1$ and $n_\infty=100\,\mu$m$^{-2}$ for the three models with the same color code.}
\label{simulations}
\end{figure}

We show in Fig.\,\ref{simulations}(a) how the ground state atom number varies with respect to $N_T$ when increasing the depletion for these three models. We also show the analytical prediction of Eq.\,\eqref{eq:Rosanov} that we rewrite as
\begin{equation}
N / N_T = 1 + 0.28 \, \epsilon. \label{eq:Rosanov_2}
\end{equation} 
We have introduced the depletion parameter
\begin{equation}
\epsilon = \frac{N_T / \sigma^2}{n_\infty},
\end{equation}
with $\sigma$ the rms size of the corresponding state, which is related to the length $\ell$ as $\sigma \simeq 1.09\,\ell$ for the Townes profile of Eq.\,\eqref{eq:townes_family}. This quantity can be viewed as the ratio between the typical peak density $N_T / \sigma^2$ in the impurity component $|2\rangle$ and the atom density $n_\infty$ in the bath component $|1\rangle$. All models predict a similar shift of the ground state atom number for values of $\epsilon \lesssim 0.25$, which is the maximum value of all data presented in the main text. We also note that the single component effective model gives a faithful description of the two-component system for both the ground state atom number and the density profile showed in Fig.\,\ref{simulations}b.

It is interesting to note that our work at small and intermediate depletions connects in the limit of full-depletion of the bath ($n \rightarrow n_\infty$ at the center of the bubble) to the physics of spin domains in an immiscible mixture, a situation in which the single-component effective equation introduced in this work may be of interest.


\section{4 - Experimental determination of the rms size}

The results presented in the main text exploit the measured rms size $\sigma^2$ defined as
\begin{equation}
\label{rms_size}
\sigma^2 = \frac{1}{N} \int  \mathrm{d}^2r\, n(\textbf{r}) \, r^2-\langle \textbf{r} \rangle^2,
\end{equation}
where $n$ is the atomic density in state $|2\rangle$.  Direct determination of the rms size is challenging experimentally. Indeed, the contribution of the points at large $r$ is important for a 2D integral and our signal to noise ratio is poor in this region. Consequently, we use a fit to the data to determine the rms size.  We detail below the choice of the fitting function and the fitting procedure.  We confirmed the validity of this method by applying it to the results of numerical simulations of the two-component NLSEs.

\vskip5pt

\paragraph{Determination of the fitting function.}
We use time-dependent perturbation theory to extract a suitable fitting function for the deformation of the density profile. We consider the evolution of a wave function $\phi$ under the time-dependent NLSE
\begin{equation}
\mi \frac{\partial \phi}{\partial \tau} = - \frac{1}{2} \nabla^2 \phi - G |\phi|^2 \phi,
\label{eq:nlse_time}
\end{equation}
with $\tau = (m / \hbar) t$. From Section\,2, we know that for $G = G_T$, the stationary solution of Eq.\,\eqref{eq:nlse_time} with chemical potential $\tilde \mu < 0$ is given by
\begin{equation}
\phi(r, \tau) = \phi_\ell(r) e^{- \mi \tilde\mu \tau},
\end{equation}
with $\ell  = 1/\sqrt{|\tilde \mu|}$. We consider a wave function $\phi$ given by a Townes profile $\phi_\ell(r)$ at $\tau = 0$, with an interaction parameter $G$ that is slightly different from $G_T$. We define the small parameter of the expansion $\eta$ such that $G = (1+\eta) G_T$. At short times, the deformation of the wave function with respect to the Townes profile is expected to be small, and we can expand the solution with respect to $\eta$:
\begin{equation}
\phi(r, \tau) = [\phi_\ell(r)  + \eta \epsilon(r, \tau) + \ldots]e^{- \mi \tilde\mu \tau}.
\end{equation}
We restrict here to the first-order correction in $\eta$, and consider the first terms of the Taylor expansion of $\epsilon(r, \tau)$ with respect to $\tau$:
\begin{equation}
\epsilon(r, \tau) = \epsilon_0(r) + \epsilon_1(r) \tau + \epsilon_2(r) \tau^2 +  \ldots \label{eq:def_epsilon}
\end{equation}
The initial condition gives directly $\epsilon_0(r) = 0$, and by injecting the expansion given in Eq.\,\eqref{eq:def_epsilon} in Eq.\,\eqref{eq:nlse_time}, we identify
\begin{equation}
\begin{aligned}
\epsilon_1(r) &= \mi \phi_\ell^3(r) \\
\epsilon_2(r) &= \frac{G_T}{2} \left( -\tilde\mu - \frac{1}{2}\nabla^2 - G_T \phi_\ell^2 \right) \phi_\ell^3.
\end{aligned}
\end{equation}
Interestingly, this last identity can be further simplified using Eq.\,\eqref{eq:defR} and we obtain
\begin{equation}
\epsilon_2(r) = G_T \left(G_T \phi_\ell^5 + \tilde \mu \phi_\ell^3 - \frac{3}{2}\phi_\ell {\phi_\ell'}^2 \right). \label{eq:formula_epsilon}
\end{equation}
Related approches were introduced in Refs.\,\cite{Zakharov68, Malkin91}. More precisely, the authors of Ref.\,\cite{Malkin91} studied the elementary excitations of the NLSE given in Eq.\,\eqref{eq:nlse_time} and looked for exact solutions that were at most polynomials on $t$, while here we do not impose such a constraint but restrict to a short time expansion.

\begin{figure}
\centering
\includegraphics[width=7.8cm]{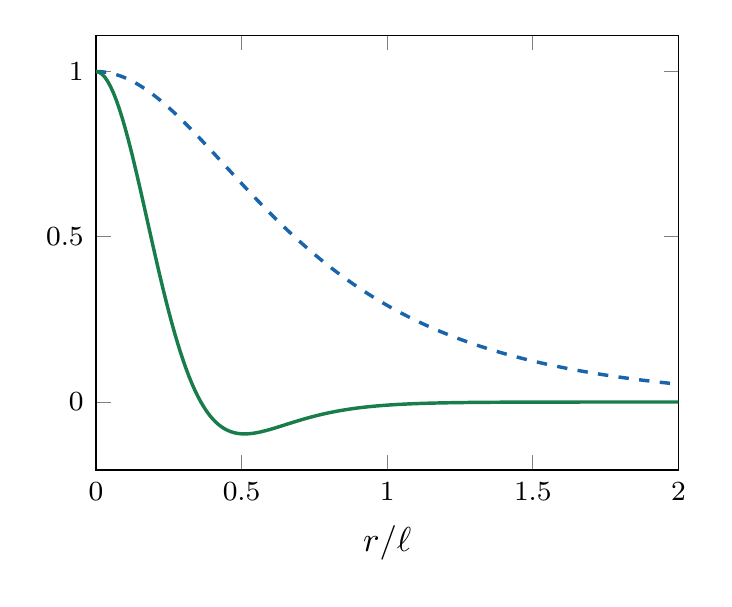}
\caption{Townes profile $\phi_\ell(r) / \phi_\ell(0)$ (blue dashed line) and $\chi(r) / \chi(0)$ (green solid line) deduced from perturbation theory and expressed in Eq.\,\eqref{eq:def_chi}.}
\label{plot_chi}
\end{figure}

When computing the density profile $n(r,t) = |\phi(r,t)|^2$, only the real term $\epsilon_2$ contributes at first order in $\eta$ (the imaginary term $\epsilon_1$ contributes to the phase of the wavefunction). We deduce the expected deformation of the density profile at first order and at short times 
\begin{eqnarray}
\delta n(r, t)&=& n(r, t) - n(r,0) \nonumber \\ 
&\simeq& 2 \eta \phi_\ell(r) \epsilon_2(r) t^2 \equiv \eta  \chi(r) t^2,\label{eq:def_chi}
\end{eqnarray}
where we have defined $\chi(r) = 2 \phi_\ell(r) \epsilon_2(r)$. We checked that the 2D integral of $\chi$ is zero, as the norm of $\phi$ should be conserved by the evolution under Eq.\,\eqref{eq:nlse_time}. In Fig.\,\ref{plot_chi} we show the profiles $\phi_\ell(r)$ and $\chi(r)$.


\vskip5pt

\paragraph{Fitting procedure.}
The first step of the data analysis consists in obtaining averaged 1D radial density profiles $n(r, t)$. The averaging is performed after recentering the individual images. Indeed, we observe random drifts of the wave packet from one shot to another, which we attribute to thermal fluctuations.

In a second step, we fit the initial profile to a Townes density profile with a free amplitude and size, which we denote $n^0(r)$. For each time of the evolution  we compute the deformation of the density profile with respect to the fitted initial one
\begin{equation}
\label{def_diff}
\delta n(r, t) = n(r, t) - \beta n^0(r),
\end{equation}
where $\beta$ is a correction factor to make the two terms of the right-hand-side of Eq.\,\eqref{def_diff} have the same atom number. 

\begin{figure}
\centering
\includegraphics[width=8.4cm]{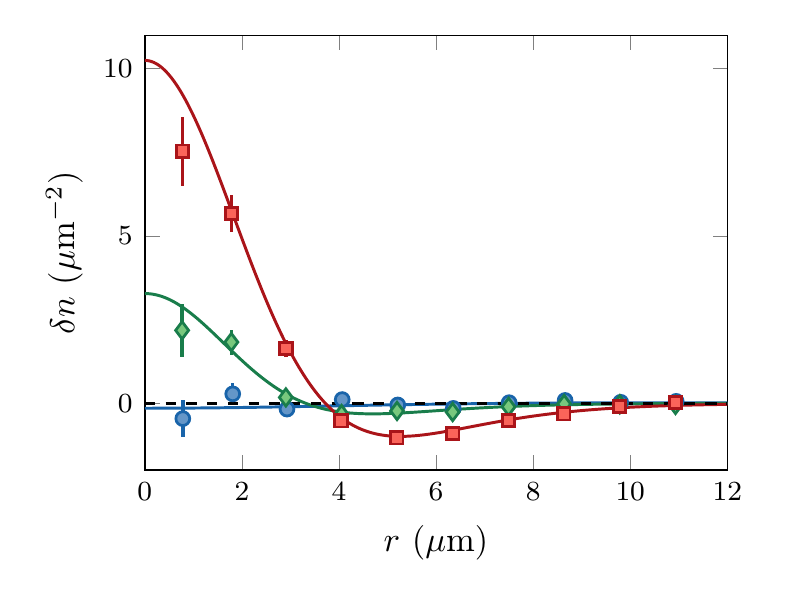}
\caption{Difference $\delta n(r, t)$ as defined in equation \eqref{def_diff}, for various times $t$ of the experimental run presented in Fig.\,2 of the main text ($N = 1200$): blue circles for $t=0.8$\,ms, green diamonds for $t = 20$\,ms, red squares for $t = 40$\,ms. Simultaneously, we plot the best fit of $\chi(r)$ to the data.}
\label{figure_fits}
\end{figure}

The last step consists in fitting this profile with the function $\chi(r)$ determined in Eq.\,\eqref{eq:def_chi} with a free amplitude and size. This fit is performed on a radial region that extends from 0 to 1.75 $\sigma_0$, with $\sigma_0$ the initial rms size (obtained from the Townes fit). Examples of such fits are reported in Fig.\,\ref{figure_fits}. We compute $\sigma$ using this fitting function over the full plane. Additionally, we estimate the error on $\sigma$ by performing a bootstrap analysis.


\section{5 - Control of the critical atom number}

\begin{figure}
\centering
\includegraphics[width=8.4cm]{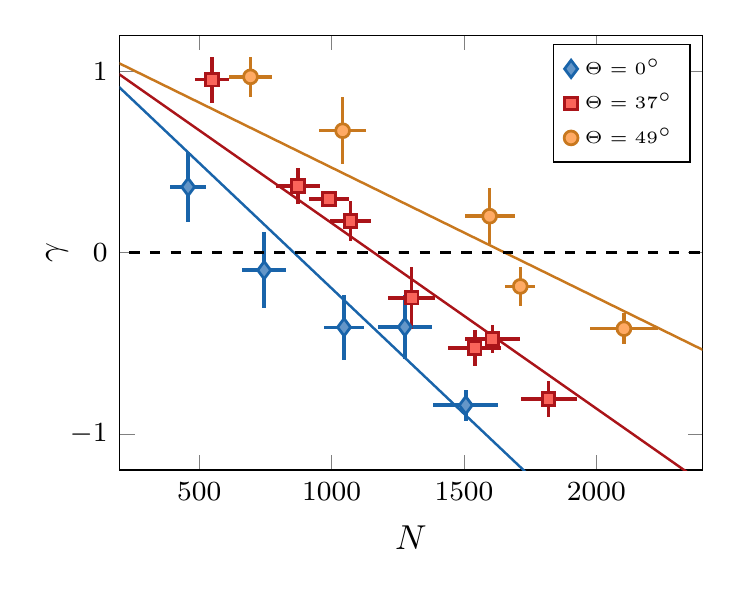}
\caption{Expansion coefficient $\gamma$ as a function of the atom number $N$ for varied orientations $\Theta$ of the magnetic field. The rms size of the imprinted cloud is set to $\sigma = 8.6\,\mu$m for all data considered here, and the bath density is $n_{\infty} = 90$ atom/$\mu$m$^2$. For each set of points we also plot the linear fit of $\gamma(N)$ from which we deduce $N_T^{\textrm{exp}}$.}
\label{gamma_theta}
\end{figure}

We studied in Ref.\,\cite{Zou20a} the dependence of $a_{12}$ with the orientation of the quantization axis given by the magnetic field $\textbf{B}$. More precisely, if we denote $\Theta$ the angle between $\textbf{B}$ and the vertical (strongly confining-)axis $\hat{z}$, we can model the 2D inter-component interactions with a dimensionless parameter $\tilde{g}_{12} = \sqrt{8\pi} a_{12} / \ell_z$ where the effective scattering-length has to be corrected from the bare (3D-)value $a_{12}^0 = 98.9\,a_0$, such that
\begin{equation}
a_{12} = a_{12}^0 + \delta a_{12}~~~ \delta a_{12} = a_{dd} \left(3 \cos^2\Theta - 1  \right),
\end{equation}
where $a_{dd} = \mu_0 \mu_B^2 m /( 12 \pi \hbar^2) = 0.7 \, a_0$ is the dipole length. Despite the smallness of this shift compared to $a_{12}^0$, it has a strong influence on the effective critical atom number $N_T = G_T/|\tilde g_e|$, which varies from $N_T(\Theta = 0^{\circ}) \sim 750$ to $N_T(\Theta = 90^{\circ}) \sim 5000$ with our experimental parameters. In Fig.\,\ref{gamma_theta}, we report our measurements of the expansion coefficient $\gamma(N)$ for different orientations $\Theta$ of the magnetic field. We restrict ourselves to  $N<2200$ to ensure the bath stays in the weak depletion limit for the sizes $\sigma \lesssim 9\,\mu$m imposed by the geometry of the experiment. From a linear fit of $\gamma$, with $\gamma(N=0)=1.19$ fixed at the expected value, we deduce the stationary atom number  $N_T^{\textrm{exp}}(\Theta)$ at which this expansion coefficient vanishes, which is shown in the inset of Fig.\,3 of the main text.


Anisotropic effects due to magnetic dipole-dipole interactions are not expected to modify the properties of the system as long as  $\sigma\gg \ell_z$, where $\ell_z$ is the vertical confinement length \cite{Zou20a}. We checked that the modification of $N_T$ should remain smaller than 5\% for all data presented here.

\section{6 - Universal properties of 2D attractive bosons}
\label{sec:6}

In Ref.\,\cite{Hammer04}, the authors studied the ground state properties of weakly interacting bosons in two dimensions using a classical field formalism with a regularized contact potential. In the following, we recall their main results and show that the expected corrections in our experimental situation are not observable.

One considers bosons in two dimensions interacting via an attractive contact potential $(\hbar^2/m) \tilde g \, \delta (\textbf{r})$, with a dimensionless coupling constant $\tilde g < 0$. The quantum treatment of the collisions is mathematically ill-defined for such a contact potential. For $|\tilde g| \ll 1$, a more accurate description of the system can be obtained by substituting the bare parameter $\tilde g$ by a running coupling constant defined by 
\begin{equation}
\frac{1}{\tilde g (k)} =\frac{1}{\tilde g} + \frac{1}{2\pi} \ln \left( \frac{k_c}{k} \right) , \label{eq:running_g}
\end{equation}
which depends on the relative momentum $k$ of the two particles involved in the collision. The introduction of a cut-off in momentum space $k_c$ is a signature of an intrinsic length scale $1/k_c$ of the physical system given by the van der Waals length scale $R_{\textrm{vdW}} \approx 5$\,nm for $^{87}$Rb. The ground state properties of the system are derived using a variational approach. Here, one considers trial wave functions with a Townes profile of extension $\ell$. The energy per particle of the classical field with $N$ atoms then writes
\begin{equation}
E_N(\ell) \propto \frac{1}{\ell^2} + C \frac{\tilde g(\ell^{-1}) N}{\ell^2}, \label{eq:energy_hs}
\end{equation}
where $C>0$ is a  numerical factor and $\tilde g(k)$ is evaluated at the typical momentum $\ell^{-1}$.

In contrast to Eq.\,(2) of the main text, $E_N$ has now a non trivial dependence on $\ell$ because of the non-constant parameter $\tilde g(\ell^{-1})$.  This term breaks scale invariance and gives rise to an equilibrium size and a binding energy $(\ell_N, E_N)$ that follow a geometrical law
\begin{eqnarray}
\ell_{N+1} \sim 0.34 \,\ell_N \qquad
E_{N+1} \sim \frac{1}{(0.34)^2} E_N. \label{eq:universal}
\end{eqnarray}
Note that $\ell_N$ and $E_N$ vary extremely rapidly with $N$. For example, one can rewrite $\ell_N$ as
\begin{eqnarray}
\ell_N \sim R_{\textrm{vdW}} \exp \left[- \zeta (N - N_T) \right],
\end{eqnarray}
with $\zeta \approx 1$. For $N=N_T =G_T /| \tilde g|$, this size is $\sim R_{\textrm{vdW}}$, which is 3 orders of magnitude smaller than the size of our system. A small shift of only a few atoms, typically from $N_T$ to $N_* \equiv N_T - 6$, gives a size of $\approx$ few microns, compatible with the extension of our system. Experimentally, we cannot resolve the difference between these two atom numbers, as it would require single-particle resolution. Going further away from $N_T$, the corresponding sizes are either much too large or much too small to be experimentally relevant. For this reason, we do not expect to observe a stable state for atom numbers differing significantly from $N_T$ on our experiment.

Finally, we remark that the breakdown of scale-invariance close to $N_*$ is too weak to be observed with our experimental setup. Indeed, consider a system with $N = N_*$ atoms. At equilibrium, Hammer \& Son \cite{Hammer04} predict an energy per particle  $E_{N_*}(\ell_*) \sim -\hbar^2/(N_*m\ell_*^2)$, which is $1/N_*$ smaller than the usual energy associated with the length scale $\ell_* \equiv \ell_{N_*}$. Therefore, if the system is prepared in a Townes profile of size $\ell$ slightly differing from $\ell_*$, the typical energy scale governing the dynamics is
\begin{equation}
E_{N_*}(\ell) - E_{N_*}(\ell_*) \sim \frac{1}{N_T}  \frac{\Delta \ell}{\ell_*} \frac{\hbar^2}{m \ell_*^2}, \label{eq:energy_diff}
\end{equation}
with $\Delta \ell = \ell - \ell_*$. This energy difference $\propto 1/N_T=|\tilde g|/G_T$ is thus negligible for $\tilde g \ll 1$ and the typical time scale considered in this work.

\end{document}